\documentclass[aps,prb,superscriptaddress,reprint]{revtex4-1}
\usepackage{graphicx}
\usepackage{latexsym}
\usepackage{amsmath}
\usepackage{amssymb}
\usepackage[permil]{overpic}
\usepackage[colorlinks]{hyperref}
\hypersetup{allcolors=blue}
\begin{document}
\title{Size-dependent thermal stability and optical properties of ultra-small nanodiamonds synthesized under high pressure}

% Authors, for the paper (add full first names)
\author{E.A. Ekimov} 
\email{ekimov@hppi.troitsk.ru}
\affiliation{Vereshchagin Institute for High Pressure Physics, Russian Academy of Sciences, Troitsk, Moscow 108840, Russian Federation}
\author{A.A. Shiryaev} 
\affiliation{Frumkin Institute of Physical Chemistry and Electrochemistry, Russian Academy of Sciences, Moscow 119071, Russian Federation}
\author{Yu. Grigoriev} 
\affiliation{Shubnikov Institute of Crystallography of Federal Scientific Research Centre, Crystallography and Photonics', Russian Academy of Sciences, Moscow, 119333, Russian Federation}
\author{A. Averin}
\affiliation{Frumkin Institute of Physical Chemistry and Electrochemistry, Russian Academy of Sciences, Moscow 119071, Russian Federation} 
\author{E. Shagieva}
\affiliation{Institute of Physics of the Czech Academy of Sciences, Cukrovarnick\'{a} 10, 16200 Prague, Czech Republic} 
\author{S. Stehlik} 
\affiliation{Institute of Physics of the Czech Academy of Sciences, Cukrovarnick\'{a} 10, 16200 Prague, Czech Republic} 
\author{M.V. Kondrin}
\email{mkondrin@hppi.troitsk.ru}
\affiliation{Vereshchagin Institute for High Pressure Physics, Russian Academy of Sciences, Troitsk, Moscow 108840, Russian Federation}

\begin{abstract}
 Diamond properties down to the quantum-size region are still poorly understood. High-pressure high-temperature (HPHT) synthesis from chloroadamantane molecules allows precise control of nanodiamond size. Thermal stability and optical properties of nanodiamonds with sizes spanning range from  $<$1 to 8 nm are investigated. It is shown that hypothesis about enhanced thermal stability of nanodiamonds smaller than 2 nm is incorrect. The most striking feature in IR absorption of these samples is the appearance of an enhanced transmission band near the diamond Raman mode (1332 cm$^{-1}$). Following previously proposed explanation, we attribute this phenomenon to the Fano-effect caused by resonance of diamond Raman mode with continuum of conductive surface states. We assume that these surface states may be formed by reconstruction of broken bonds on the nanodiamond surfaces. This effect is also responsible for the observed asymmetry of Raman scattering peak. he mechanism of nanodiamond formation in HPHT synthesis is proposed, explaining pecularities of their structure and properties.
\end{abstract}

\maketitle

\section{Introduction}

Current methods of mass production of nanodiamonds (detonation method, chemical vapor deposition, milling of high pressure high temperature (HPHT) microcrystals) do not allow simultaneous control of the size distribution of obtained samples and preservation of the structural and chemical state of the nanodiamond (ND) surfaces. These difficulties are especially pronounced in the ultra-small (or quantum) size range of 1-3 nm and thus significantly hinder study and understanding of the size dependence of the properties of NDs \cite{stehlik:jpcc15,stehlik:acsami17,shenderova:jvst19,ekimov:pu17,stehlik:jpcc21}. Insufficient understanding of the surface properties is detrimental for solution of an important fundamental and applied problem of size effect on the thermal stability and graphitization of nanodiamonds  \cite{kuznetsov06}. 

Some theoretical studies predict that nanodiamond grains with sizes below certain threshold value are more stable than graphite.  These sizes are quoted as <6 nm \cite{zhao:drm02}, 20 nm \cite{sun:jpcl14} or in a narrow size range of 1.9-5.2 nm \cite{barnard:jcp03}. Most of experimental attempts to address this issue employ detonation nanodiamond (DND). For example, based on surface graphitization rate Wang {\em et al.} \cite{wang:acie05} suggested an increase in the thermal stability of NDs with size less than 3 nm compared to larger one, 5-6 nm.  On the contrary, it was reported by Butenko {\em et al.} \cite{butenko:jap00} and Xu {\em et al.} \cite{xu:drm02} that small nanoparticles (< 2 nm) graphitize more quickly than larger ones due to higher concentration of the surface defects and larger surface energy. In many cases, conclusions about the size-dependence of thermal stability are based on structural analysis of the graphitization products in assumption their sizes are directly related to dimensions of starting ND grains \cite{qiao:sm06,kuznetsov05, tomita:jcp01}, but this statement is not necessarily always correct.  For example, during graphitization of sufficiently large submicron diamonds,formation of nanoscale graphitic structures (carbon dots) with dimensions comparable with nanodiamonds is observed \cite{kuznetsov:jap99}. In addition, the thermal stability of NDs may depend on the state of the surface. Subsequently, size-dependent properties of DND grains may vary as a function of treatment history \cite{kuznetsov05,kuznetsov06}. 

Recent breakthrough in nanodiamond synthesis in halogenated hydrocarbon systems at high static pressures (HPHT) enabled precise control of the ND size and made possible growth of nanodiamonds with desired dimensions starting from values as small as 1 nm  \cite{davydov:jetpl14,ekimov:c19, ekimov:drm20, ekimov:mrb21, kondrin:cnm21}. This provide an excellent opportunity for refinement of thermal, structural, optical, and other properties of diamond on the ultra-small scale, which is important for the understanding and revealing their prospects in future quantum and biomedical technologies. In the current paper we present study of the size-dependent thermal stability/graphitization and optical properties of nanodiamonds synthesized in the growth medium comprising C-H-Cl. A distinguishing feature of nanodiamonds produced by this approach is their surface hydrogenation \cite{davydov:jetpl14,ekimov:mrb21,kondrin:cnm21}. We propose that, contrary to the recently published preprint \cite{kudryavtsev21}, the surface of the ultra-small nanodiamonds grown from hydrocarbon precursor is largely H-free, is reconstructed and consists of contiguous network of sp$^2$ hybridized carbons with patches of sp$^3$ hydrogenated surface carbon atoms. 

%%%%%%%%%%%%%%%%%%%%%%%%%%%%%%%%%%%%%%%%%%
\section{Materials and Methods}

Synthesis of samples was carried out in a  toroid-type high-pressure chamber at pressures 8-9 GPa and temperatures up to 1300 $^\circ$C. C$_{10}$H$_{15}$Cl wax-like compound (Sigma Aldrich, 98\%) was pressed into semitransparent tablet of 60 mg and placed inside a graphite crucible heater with inner diameter 4 mm.  In each experiment the temperature control was ensured by chromel-alumel thermocouple, whose junction was fixed outside of the graphite heater 4 mm in height and 6 mm in outer diameter. Graphite mushroom-like electrical leads with ZrO$_2$ bushings were mounted at the ends of the crucible and the heating rate to desirable temperature was less than 7 K/s and characteristic dwell time at constant P-T parameters was 120 s. Pressure in the high-pressure cell was estimated by means of previously constructed calibration plot of the pressure inside chamber as a function of the oil pressure inside the hydraulic system of the press. Pressure calibration of the chamber was performed at room temperature using phase transitions in bismuth \cite{kondrina:drm18}. 

X-ray diffraction patterns were obtained in a reflection mode with use of Cuk$\alpha$ radiation and silicon sample holder. {\em In situ} high temperature diffraction experiment was performed using HTK-1200N oven in vacuum (10$^{-4}$ mbar at the highest temperature and down to 10$^{-6}$ mbar at lower temperatures). The maximum temperature was 1150 $^\circ$C, the ramp rate 10 $^\circ$ /min. XRD patterns were recorded both immediately after reaching the preset temperature step and after 30 minutes annealing; total duration of each step was thus 60 min. Platinum foil was used as a substrate. The position of the sample on the diffractometer axis was adjusted automatically.

Detonation nanodiamond (DND, average diamond crystallite size 4.5 nm, C $>$ 97 \%, All-Union Research Institute of Technical Physics (VNIITF), Snezhinsk, Chelyabinsk region, Russia) was used as a reference material in situ high temperature diffraction experiments.

Raman spectra were recorded at room temperature using the 405 nm excitation laser. The 50$\times$ objective was used for laser focusing and the laser power was kept at about 0.2 mW at the sample.  Infra-red spectra were recorded using Nicolet iN10 IR microscope. Optically transparent pieces of diamond samples were placed on KBr pellet without any grinding and studied in transmission geometry with apertures between 50 and 150 microns to match size of the sample. Spectral resolution was 2 cm$^{-1}$, up to 256 scans were recorded for each sample. Spatial homogeneity of the prepared samples was confirmed by consistent results/spectra (FTIR, Raman) obtained from several randomly chosen spots. 

TEM experiments were carried out using high-resolution transmission electron microscope FEI Tecnai Osiris with field emission gun operating on 200kV. The nanodiamond samples were mixed with acetone, dispersed in an ultrasonic bath and then deposited to a 3mm copper mesh with lacey carbon for electron microscopy. 

Electrical conductivity was measured by multimeter Agilent 34410A (Agilent, Santa Clara, USA). The samples were dispersed in toluene (Penta, Prague, Czech Republic) in concentration of 1 mg/ml by means of rod-like sonicator (Hielscher UP 200s, Teltow, Germany) working at 100\% amplitude and 50\% on/off period for one hour. Dispersed samples were dropcasted (10 $\mu$l) on an interdigitated Au electrode sensor (Micrux, Gijon, Spain) with 150 nm Au layer thickness, 10 $\mu$m electrode separation, and 90 electrode pairs. The sensor substrate material was glass.

Fourier transform infrared (FTIR) spectra of annealed samples were measured using an N$_2$-purged spectrometer (Nicolet iS50; Thermo Fisher Scientific, Waltham, MA, USA) equipped with a KBr beam splitter and N$_2$-cooled MTC HighD* detector. The spectra were measured by a specular apertured grazing angle reflectance method with an 80 $^\circ$ incident light. Each spectrum represents an average of 128 scans per spectrum collected with a resolution of 4 cm$^{-1}$. The toluene-based samples were drop-casted (50 $\mu$l) on gold-coated Si substrates, and the toluene was evaporated at 70 $^\circ$C. A bare gold-coated Si substrate was measured as a background prior to each sample measurement. 
%%%%%%%%%%%%%%%%%%%%%%%%%%%%%%%%%%%%%%%%%%
\section{Results}
\subsection{Synthesis}
\begin{figure}
\includegraphics[width=\columnwidth]{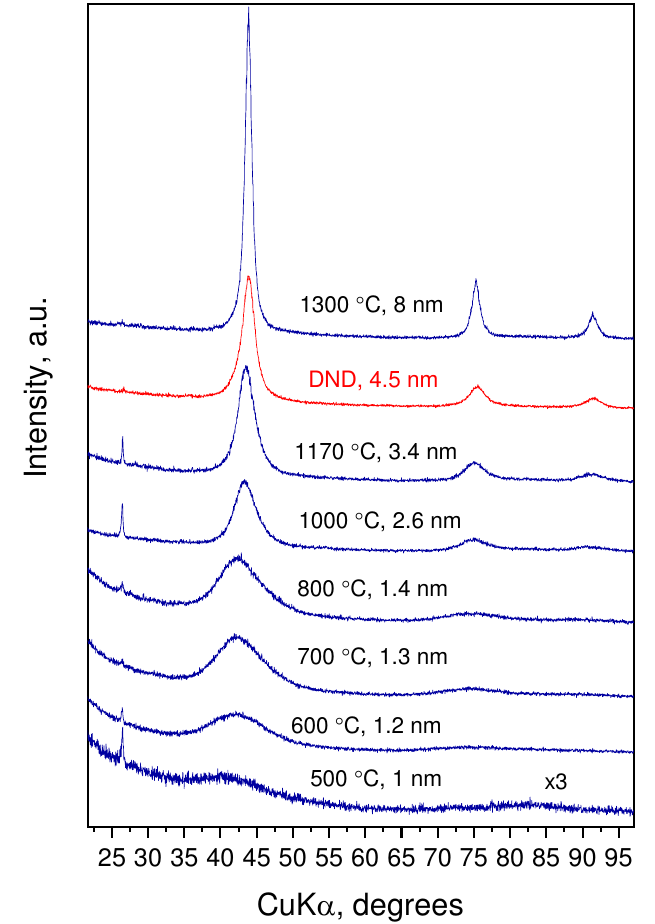}
\caption{Diffraction patterns of nanodiamond samples, obtained at 8-9 GPa and temperatures up to 1300 $^\circ$ C, and detonation nanodiamond (DND) as a reference. Synthesis temperature and average size of nanodiamonds are presented above corresponding XRD pattern.}
\label{f:2}
\end{figure}

\begin{figure*}
\includegraphics[width=\textwidth]{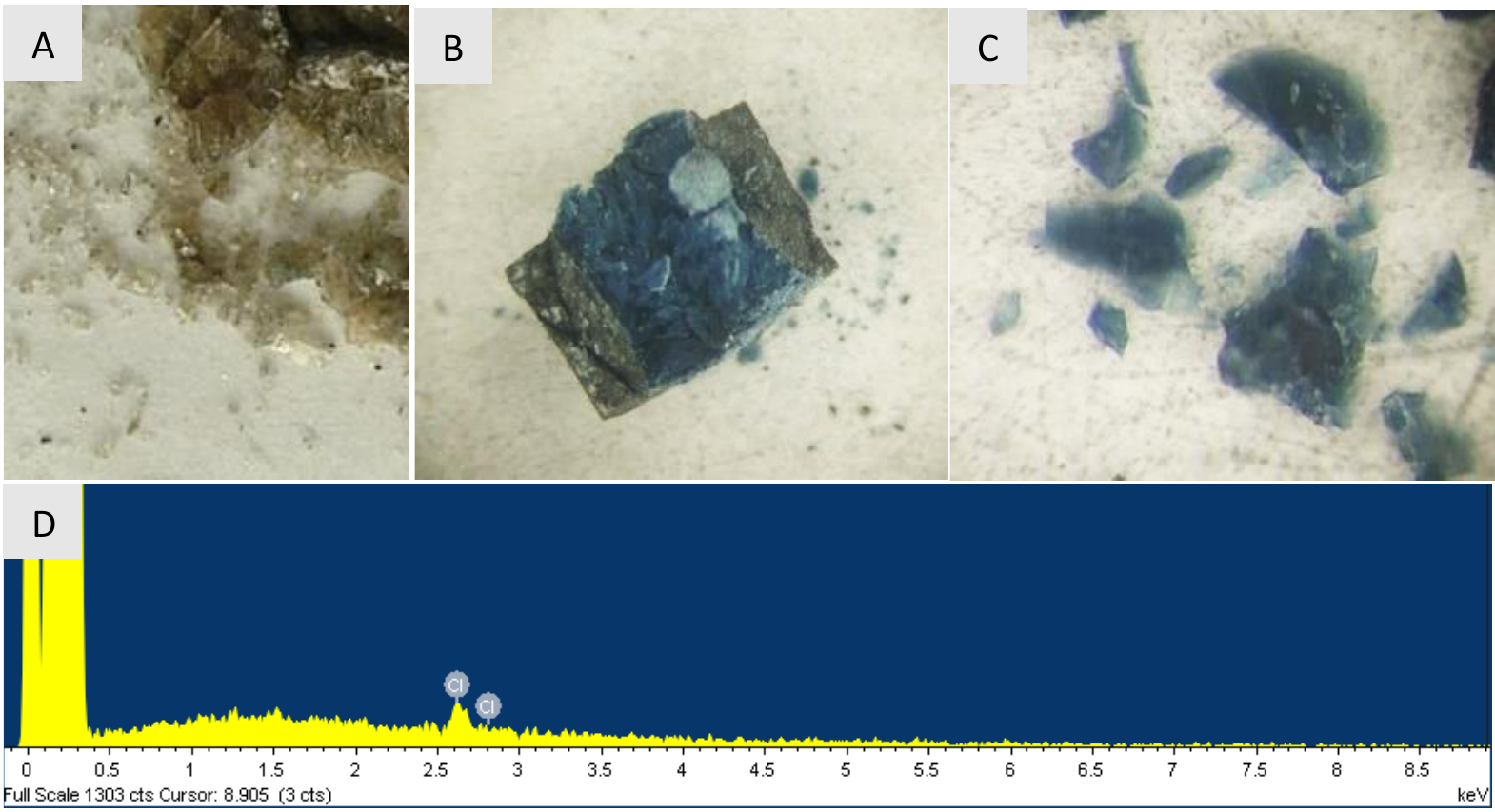}
\caption{ Pieces of nanodiamonds after synthesis at 800 $^\circ$C (A) and 1300 $^\circ$C (B, C). (B) -- nanodiamond sample in an opened graphite heater, (D) shows the characteristic X-ray  fluorescence  spectrum of the sample synthesized at 1300 $^\circ$C, demonstrating presence of only carbon and residual chlorine  ($\approx$ 0.7 at. \%). Note that hydrogen cannot not detected by this technique.}
\label{f:1}
\end{figure*}

\begin{figure}
\includegraphics[width=\columnwidth]{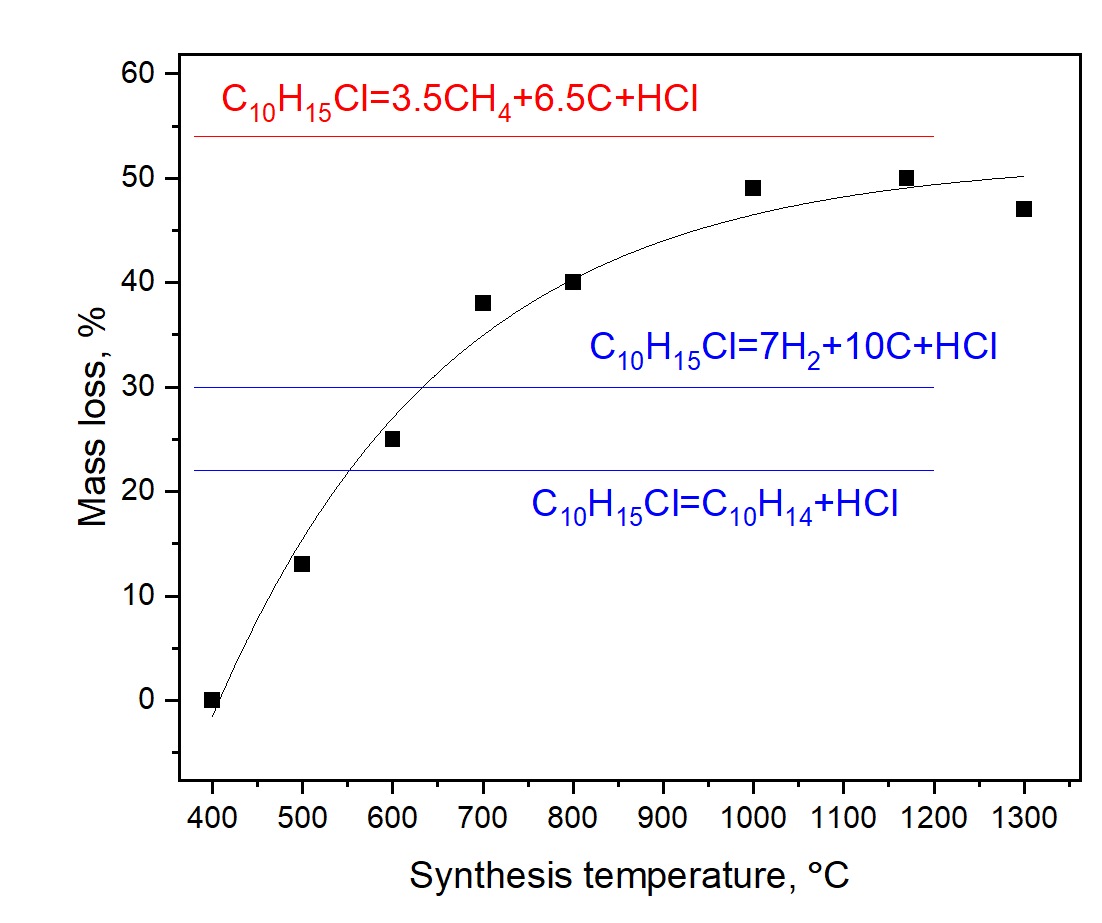}
\caption{Mass loss of C$_{10}$H$_{15}$Cl as function of the synthesis temperature. Blue labels show possible mass losses including only dehydrochlorogenation route of synthesis reaction. Red label demonstrates the actual asymptote of mass loss observed in experiments.}
\label{f:3}
\end{figure}

Decomposition of C$_{10}$H$_{15}$Cl at 8-9 GPa begins at 500 $^\circ$C and is manifested as decrease in the sample mass and changes in X-ray diffraction patterns (Fig.~\ref{f:2}). Perfect matching of diffraction peaks to diamond (111), (220), and (311) reflections and their evolution with increased synthesis temperature clearly indicates nucleation of nanodiamonds. The sharp graphite peak at 26 $^\circ$ ($2 \Theta$) is due to pieces of graphite heater from the reaction cell.Size of ND core was determined using the Scherrer equation based on the (111) reflection. Average crystal size of nanodiamonds increases from about 1 to 8 nm with increasing synthesis temperature from 500 to 1300 $^\circ$C. 

Samples obtained from C$_{10}$H$_{15}$Cl at temperatures up to 1300 $^\circ$C have a glassy appearance and are translucent. The color of the samples changes from brown to blue at synthesis temperatures above 800 $^\circ$C (see Fig.~\ref{f:1}). Apart from C and Cl, as components of the reaction system, no impurities, including  ubiquitous oxygen and nitrogen were detected in the synthesized samples by EDX analysis. In samples synthesized at  temperature of 1300 $^\circ$C EDX analysis reveals only Cl  at $<$ 0.7 mass \% level. Quantitative atomic emission spectral analysis of the samples performed using the standards did not show any presence of B, Si, Mg, Fe, Al, Ca, Cr at levels above  0.01-0.001 mass\% levels. In contrast, the studied DND powder contains Si (0.01 mass\%), Mg (0.002 mass\%), Fe (0.01 mass\%) Al (0.01 mass\%), Cr (0.06 mass\%), Ca (<0.01 mass\%) likely inherited from synthesis and purification. This documents high chemical purity of the prepared samples.

The mass loss of the samples gradually increases with the synthesis temperature and reaches saturation at temperatures above 800 $^\circ$C (Fig.~\ref{f:3}). The mass loss above this temperature corresponds to removal of hydrogen and chlorine from the system in form of CH$_4$ and HCl, although formation of free hydrogen, chlorine and other volatile components is not excluded.

Combined consideration of X-ray diffraction and mass loss data and taking into account well-studied pyrolysis of polyvinyl chloride (PVC) at normal pressures \cite{kondrin:cnm21} suggest that the decomposition of chloradamantane C$_{10}$H$_{15}$Cl at temperatures above 500 $^\circ$C  could start from dehydrochlorination, subsequent intermolecular linking, resulting in formation of single C-C bonds allowing nucleation of the of the smallest stable 1 nm nanodiamonds. However, the reaction of chloroadamantane molecule is probably more complicated in comparison to PVC, where mass loss during synthesis indicates simple dehydrohalogenation route with no loss of carbon atoms. From topological point of view the diamond structure can be represented by two interpenetrating networks of polymerized diamantane molecules. Formation of diamantane molecules may proceed via reaction between two adamantanes with subsequent partial destruction of one of them (detachment of one cyclohexane backbone involving the splitting of C-C bond). The exact mechanism of chloroadamantane polymerization requires additional studies.

We interpret the observed growth of nanodiamonds up to 8 nm at higher synthesis temperatures as the result of recrystallization of nanocrystals, the driving factor of which is a decrease in the surface energy of particles. Above 1300 $^\circ$C the graphite heater partially transforms into diamond leading to formation of crystals up to several microns in diameter. This indicates catalytic properties of a Cl-containing hydrocarbon medium in graphite-to-diamond transformation. Presumably, attempts to grow large crystals in such media require use of graphite or carbon-containing substances which are not prone to spontaneous nanodiamond nucleation. An alternative way to stimulate the growth of nanodiamonds is to carry out experiments at pressures allowing partial graphitization of chlorinated diamantane. The resulting graphite can serve as a carbon source for the further growth of diamonds.

\subsection{Infrared Spectroscopy}
\begin{figure}
\includegraphics[width=\columnwidth]{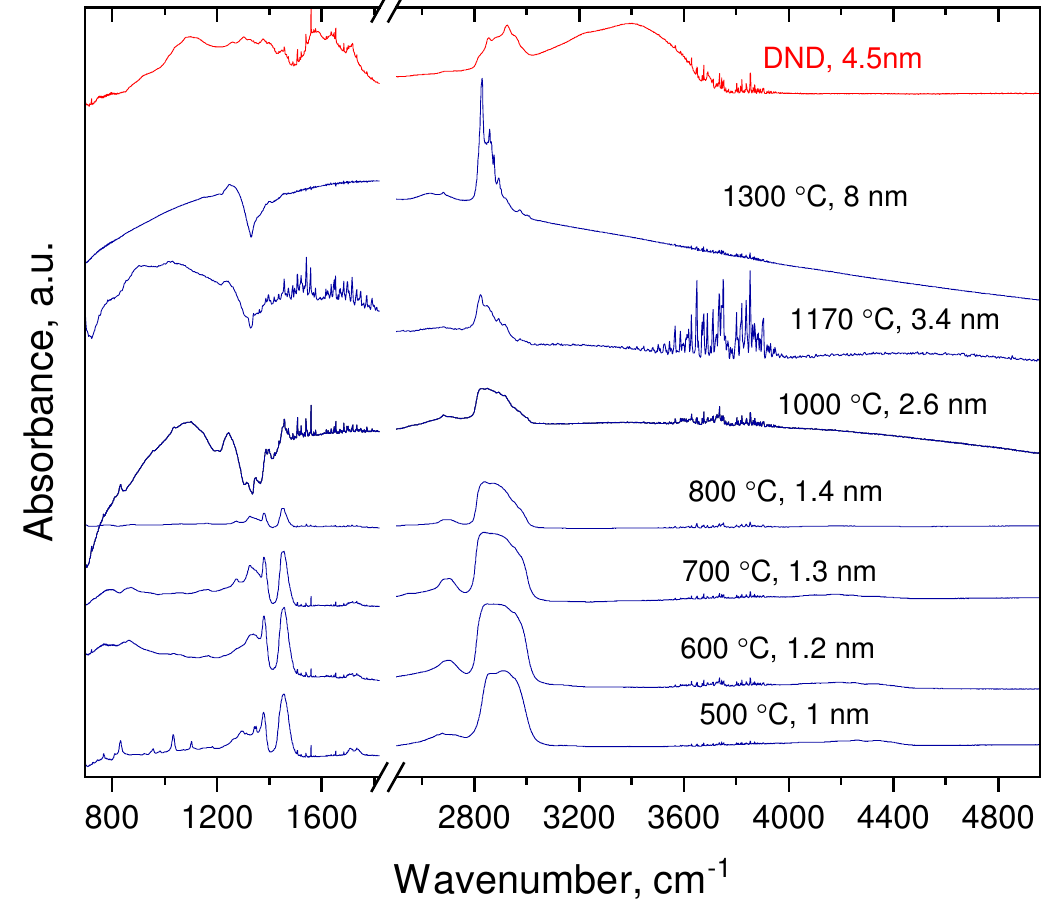}
\caption{Evolution of the IR absorbance of nanodiamonds with synthesis temperatures/average crystal size.  }
\label{f:4}
\end{figure}

FTIR spectrum of DND, recorded as a reference (Fig.~\ref{f:4}), is very similar to those published previously and consists of broad complex band around 1000 cm$^{-1}$ due to bending vibrations of various C-H and C-O groups and C-H related stretching vibrations around 2800 cm$^{-1}$. The similar features are discernible in the work of Mermoux {\em et al.} observed on DND samples annealed in vacuum \cite{mermoux:jpcc14}. Following methodology for band assignment of hydroxylated and hydrogenated nandiamonds \cite{shenderova:jpcc11,stehlik:drm16,stehlik:c21}, strong absorbance in 3000-3800 cm$^{-1}$ with features at 3230 and 3400 cm$^{-1}$ is assigned to  adsorbed water and hydroxyl groups on DND.  Band at 2800-3000 cm$^{-1}$ was attributed to symmetric and asymmetric aliphatic CH stretching vibrations of CH$_3$ (2950 and 2880 cm$^{-1}$ ) and CH$_2$ (2924 and 2850 cm$^{-1}$ ), correspondingly. Characteristic peaks at 1460 and 1375 cm$^{-1}$ were ascribed to deformation vibrations $\delta$ (CH$_2$), $\delta_{as}$ (CH$_3$)  and $\delta_s$ (CH$_3$) of -CH$_3$ and  >CH$_2$ groups, while feature at 1577, 1640 and 1720 cm$^{-1}$ to stretching $\nu$( aromatic C=C), bending $\delta$ (-O-H of adsorbed water) and (>C=O) vibrations, respectively. The 1577 cm$^{-1}$ feature can also be related to surface-confined water with limited influence of the hydrogen bonds  of hydrogenated DND \cite{petit:jpcc17}.

\begin{figure}
\includegraphics[width=\columnwidth]{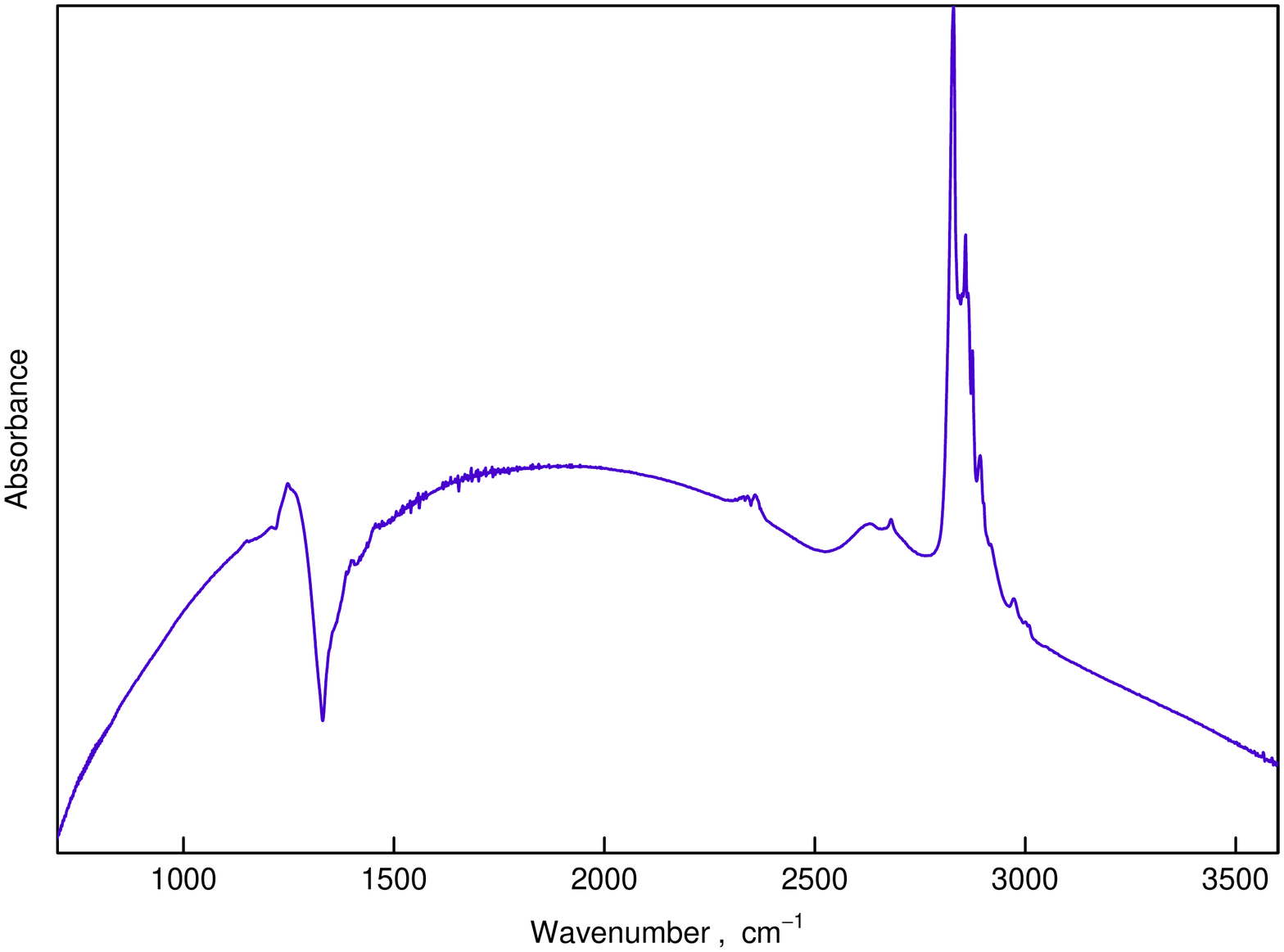}
\caption{IR absorbance of the sample synthesized at 1300 $^\circ$C (8 nm in size) in the wavenumber range 700-3600 cm$^{-1}$. }
\label{f:4a}
\end{figure}

FTIR spectra of the samples obtained by pyrolysis of C$_{10}$H$_{15}$Cl under pressure differ from those of DND (see Fig.~\ref{f:4}). In the omitted region (1800-1600 cm$^{-1}$) no prominent features was visible except C=O vibrations band at 2300 cm$^{-1}$ (due to atmospheric CO$_2$). Fig.~\ref{f4:a} shows the spectrum of the sample synthesized at 1300 $^\circ$C with well-compensated peak of atmospheric CO$_2$.  The lack of 1640 cm$^{-1}$ and 3000-3800 cm$^{-1}$ bands implies absence of adsorbed water and/or hydroxyl groups in the nanodiamond samples. The absence of the 1577 cm$^{-1}$  peak manifests also that nonhydrogen-bonded water or alkene groups are not present in the samples and the formation of aromatic and diene hydrocarbon molecules do not take place in course of the carbonization \cite{petit:jpcc17}. This conclusion is corroborated by suppression of absorbance at the 3000-3100 cm$^{-1}$ range characteristic for vibration of C-H bonds associated with double C=C bonds. In addition to strong peaks observed at $\approx$ 1460 and 1375 cm$^{-1}$ (both are bending C-H bonds, observable also in hydrogenated-DNDs) weak features at 1330 cm$^{-1}$, 1270 cm$^{-1}$ and 2970 cm$^{-1}$ appear  in spectra of samples synthesized at temperatures below 1000 $^\circ$C. These features can be attributed to C-H vibrations of carbon atoms probably bonded to heavy atoms ({\em e.g.}, Cl, {\em i.e.} CCl-H) \cite{beltran:epj97}. 

With an increase in the synthesis temperature up to 800 $^\circ$C the C-H  band at 3000-2800 cm$^{-1}$ changes:  2970 cm$^{-1}$ component and all  low frequency components become weaker relative to the 2835 cm$^{-1}$ feature. 

Starting from a synthesis temperature of 1000 $^\circ$C, the IR spectra of samples change radically: set of absorption dips in the range 1100 -1460 cm$^{-1}$ with pronounced minima at 1200, 1305, 1330, 1360 and 1410 cm$^{-1}$ appears and the 2800-3000 cm$^{-1}$ band demonstrates new structural features. The structure of the dips evolves with synthesis temperature increase and at 1300 $^\circ$C a single well-developed dip at 1330 cm$^{-1}$ is observed. Simultaneously the features in the 2800-3000 cm$^{-1}$ band become sharper.

\subsection{Raman spectroscopy}

\begin{figure*}
\includegraphics[width=\textwidth]{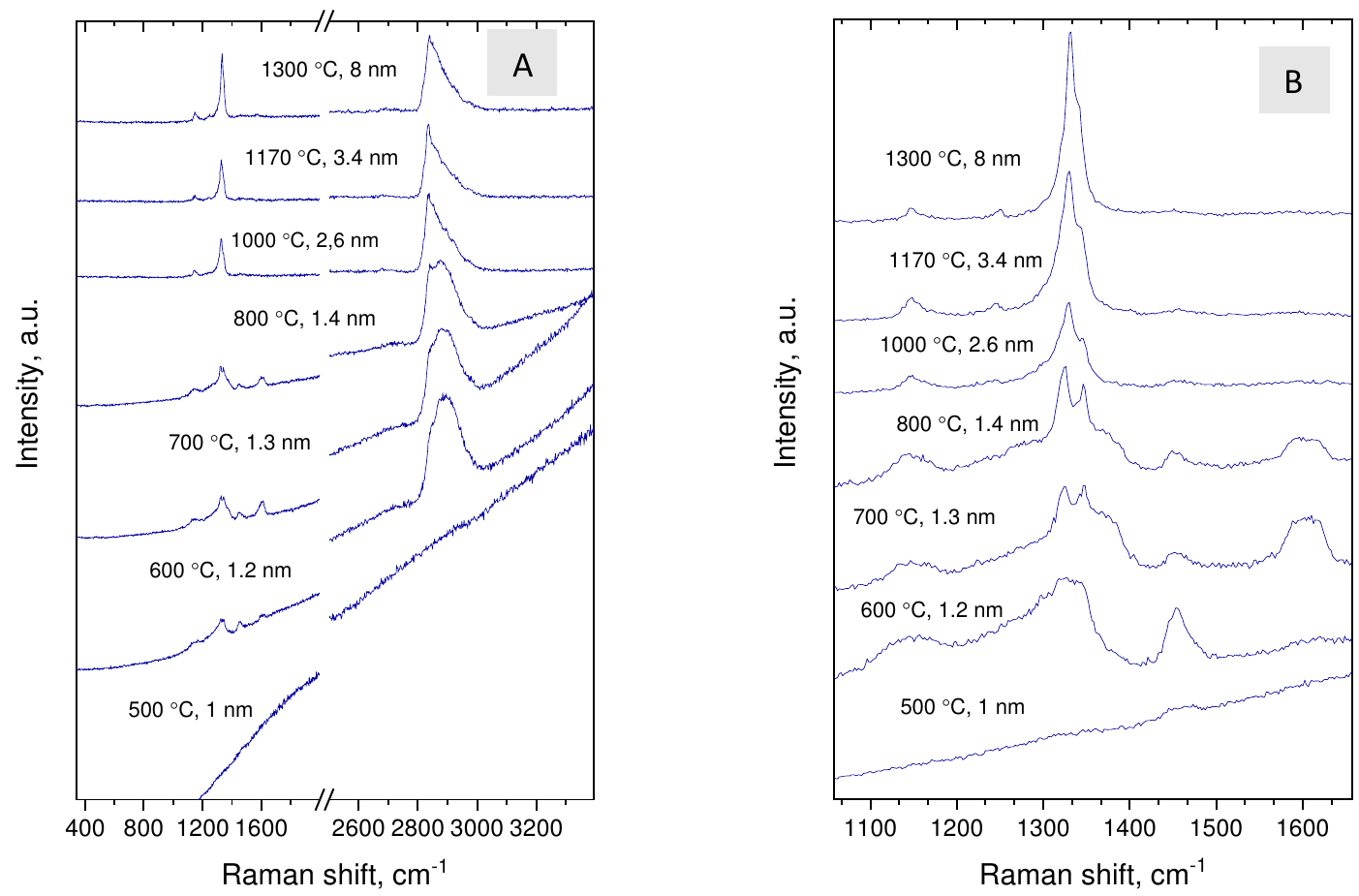}
\caption{Evolution of the Raman scattering of nanodiamonds with synthesis temperatures/average crystal size (A, B).  }
\label{f:5}
\end{figure*}

In Raman spectra of samples (Fig.~\ref{f:5}), apart from the diamond line at around 1330 cm$^{-1}$, broad band at 2800-3000 cm$^{-1}$, peaks at 1147, 1250, 1345, 1450 and broad bands at 1370 and 1560-1530 cm$^{-1}$ are observed.  Features at 1147, 1250 and 1450 cm$^{-1}$ are assigned to trans-polyacetylene fragments \cite{ferrari:ptrsa04}, and broad band in range 2800-3000 cm$^{-1}$ to C-H vibrations \cite{chang:jpc95,cheng:drm05}. The line at 1345 cm$^{-1}$ (shoulder of the diamond Raman line) is assigned to deformation mode of C-H bond \cite{sheppard:jcp48} or originates from Fermi resonance of corresponding C-H vibrations \cite{harada:saa86}. The broad bands at 1370 and 1560-1530 cm$^{-1}$, assigned as D and G lines of sp$^2$ carbon \cite{ferrari:ptrsa04}, are seen only in spectra of two samples synthesized at 700 and 800 $^\circ$C. The presence of disordered sp$^2$ carbon may reflect partial decomposition of carbonaceous precursor or be connected with stabilization of diamond structure with size less than 2 nm \cite{barnard:prb03, chang:nh18}. 

With  increasing synthesis temperature the diamond Raman line shifts from about 1322 to 1331 cm$^{-1}$ and all hydrogen-related lines, including line  at 1345 cm$^{-1}$, become relatively weak. This is in accord with increase in the size of nanodiamond crystals. At the same time, the fine structure of the 2800-3000 cm$^{-1}$ band of nanodiamonds changes drastically at synthesis temperature of about 800 $^\circ$C, which might be connected with size-dependent evolution of the crystal shape.  Domination of the line at 2835 cm$^{-1}$ in the C-H band is characteristic for diamonds with octahedral habit, C(111)-1 $\times$ 1 surface \cite{chang:jpc95}, while broad band with maximum at 2900 cm$^{-1}$ indicate absorption of C-H groups on the 110 and 100 faces \cite{cheng:drm05}. The spectroscopic data suggest that whereas nanodiamonds grown at the lowest temperatures are presumably faceted by 110 and 100 faces and may be flattened, the synthesis at higher temperatures leads to formation of predominantly octahedral grains. The transition occurs at $\approx$ 2 nm size. 

\subsection{Thermal behavior of nanodiamond samples. Graphitization.}

\begin{figure*}
\includegraphics[width=0.32\textwidth]{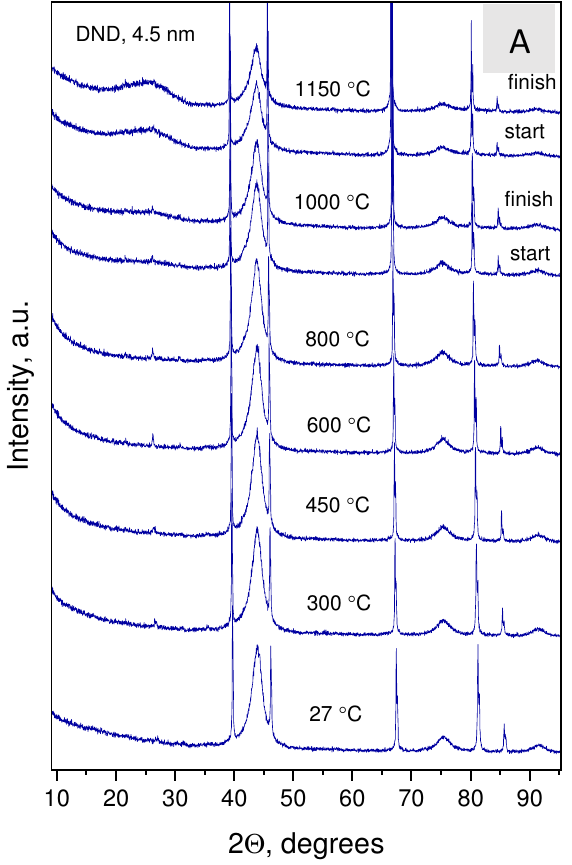}
\includegraphics[width=0.32\textwidth]{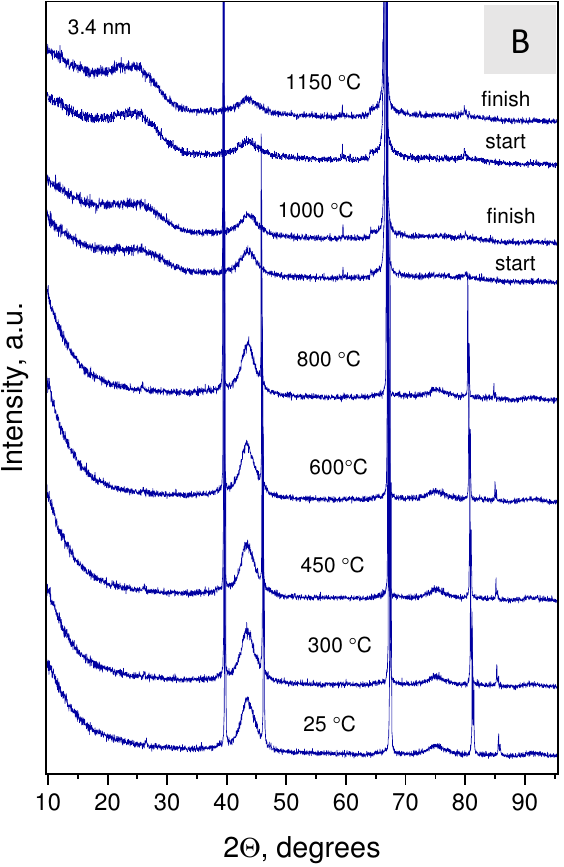}
\includegraphics[width=0.32\textwidth]{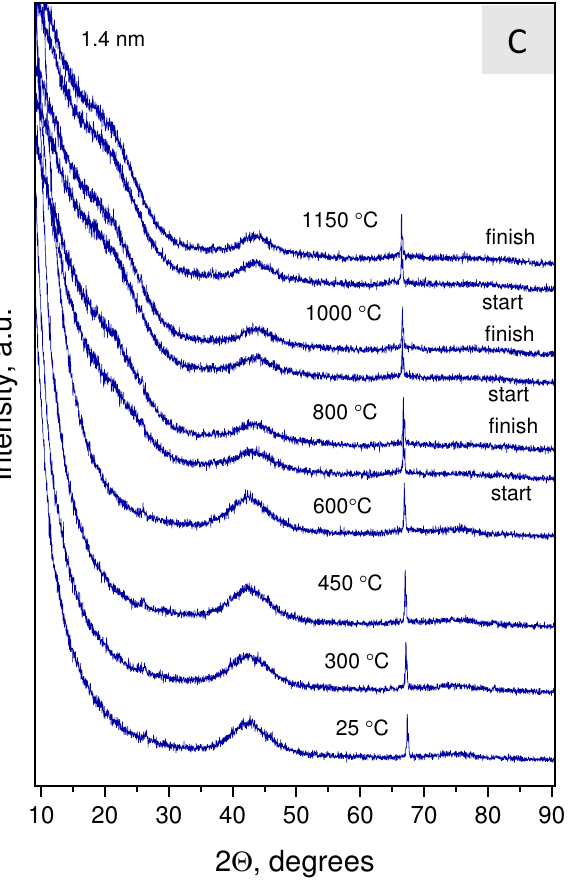}
\caption{The evolution of the phase composition traced by the XRD method during vacuum annealing of DND (4.5 nm) (A) and two nanodiamond samples synthesized at 1170 $^\circ$C (3.4 nm) (B) and 800 $^\circ$C (1.4 nm) (C). }
\label{f:6}
\end{figure*}

Fig.~\ref{f:6} shows the diffraction patterns obtained at the end of  intermediate stages of annealing where no changes relative to the original sample were observed. The presence of sharp peaks on the patterns corresponds to reflections of the platinum substrate. The temperatures for nanodiamond graphitization were determined by the appearance of a wide (002) peak of disordered graphitic carbon at $\approx$26 $^\circ$ and by a change in the profile of the (111) diamond reflection. The determined graphitization temperatures were 1000 $^\circ$C for DND and the 3.4 nm NDs and 800 $^\circ$C for the 1.4 nm NDs. Unlike DND, graphitization of the 3.4 nm NDs proceeds more intensively.

\begin{figure*}
\includegraphics[width=\textwidth]{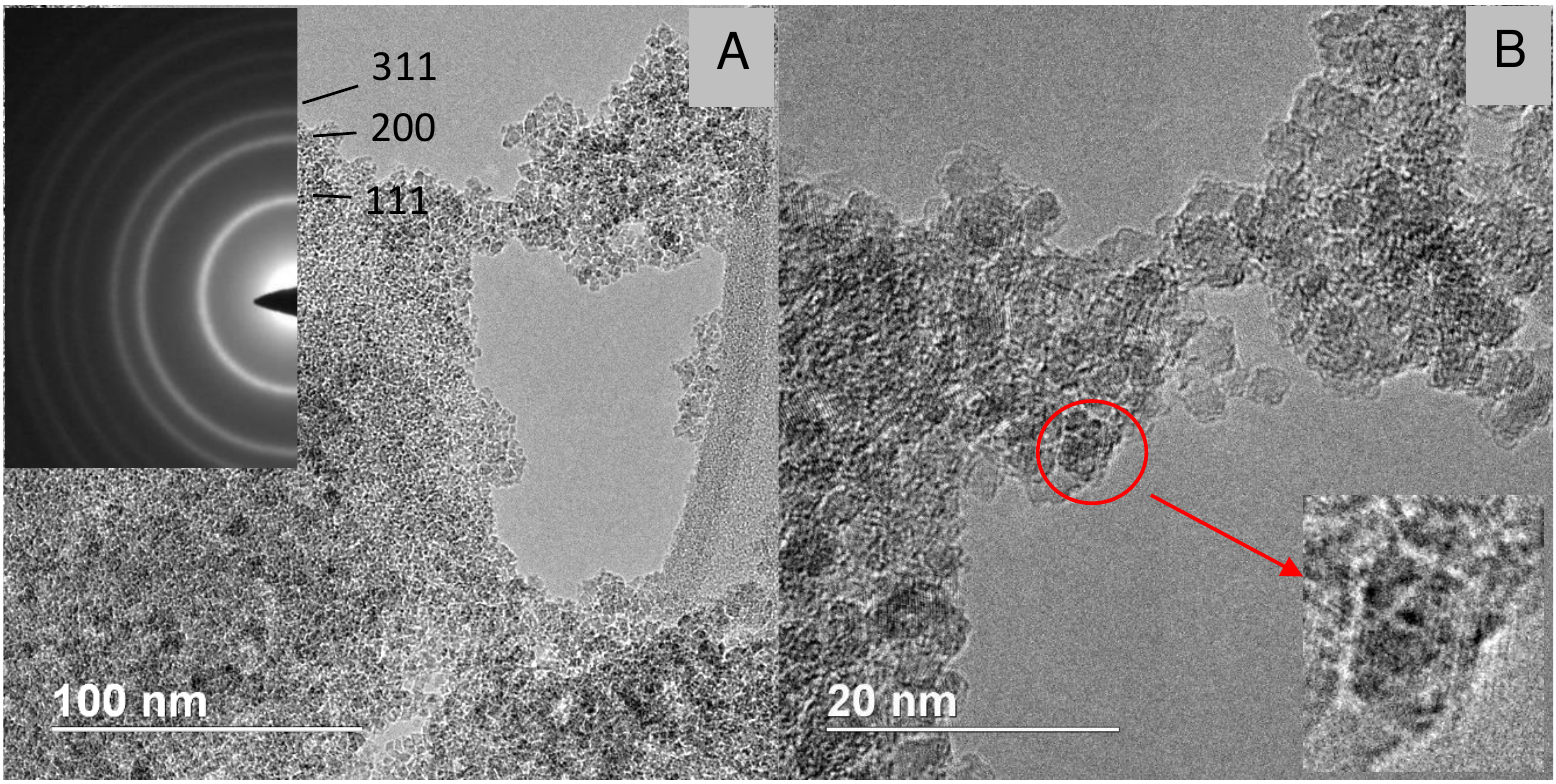}
\caption{ TEM micrographs (A, B) of the ``3.4 nm'' nanodiamonds synthesized at 1170 $^\circ$C. Insert in (A): electron diffraction pattern of the nanodiamonds. }
\label{f:7}
\end{figure*}

\begin{figure*}
\includegraphics[width=\textwidth]{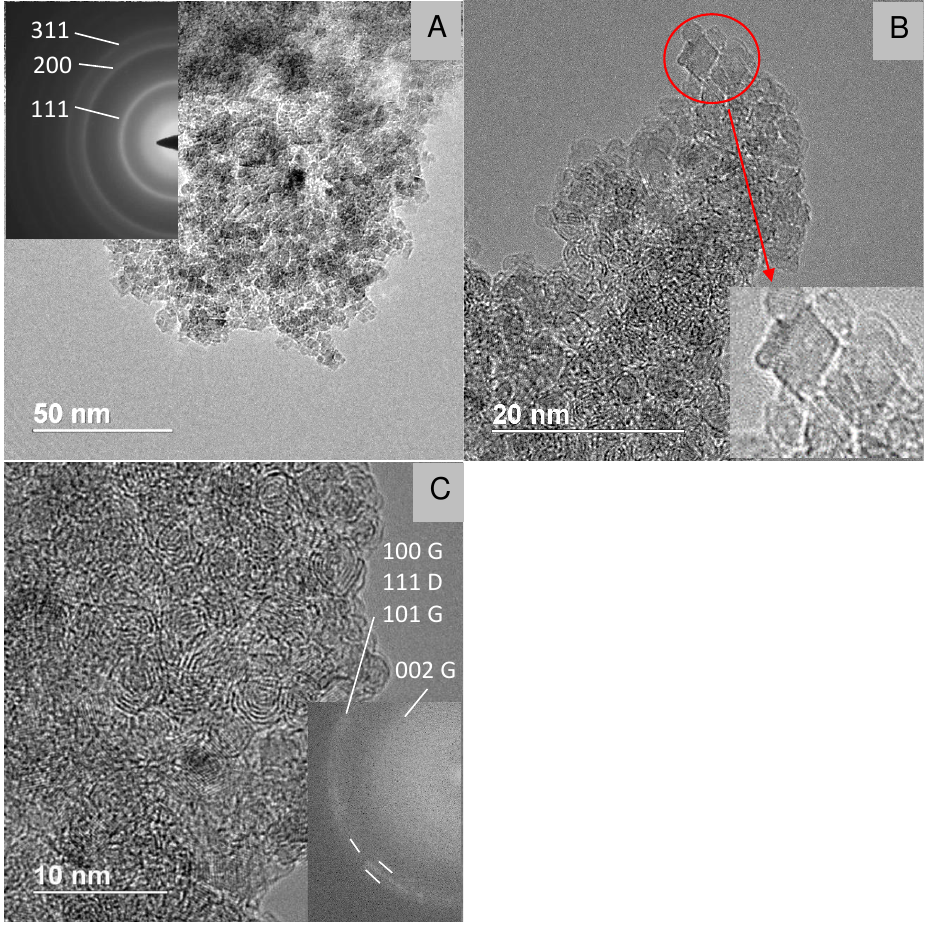}
\caption{ TEM micrographs (A-C) of the annealed ``3.4 nm'' nanodiamonds. Insert in (A): electron diffraction pattern of nanodiamonds. Insert in (C): Fourier transform calculated from the image. G -- graphite, D -- diamond. Three white strokes in the lower part of inset in panel C correspond to 100G, 111D and 101G reflections (ordered accordingly to the decrease of the diffraction angle).}
\label{f:8}
\end{figure*}

\begin{figure}
\includegraphics[width=\columnwidth]{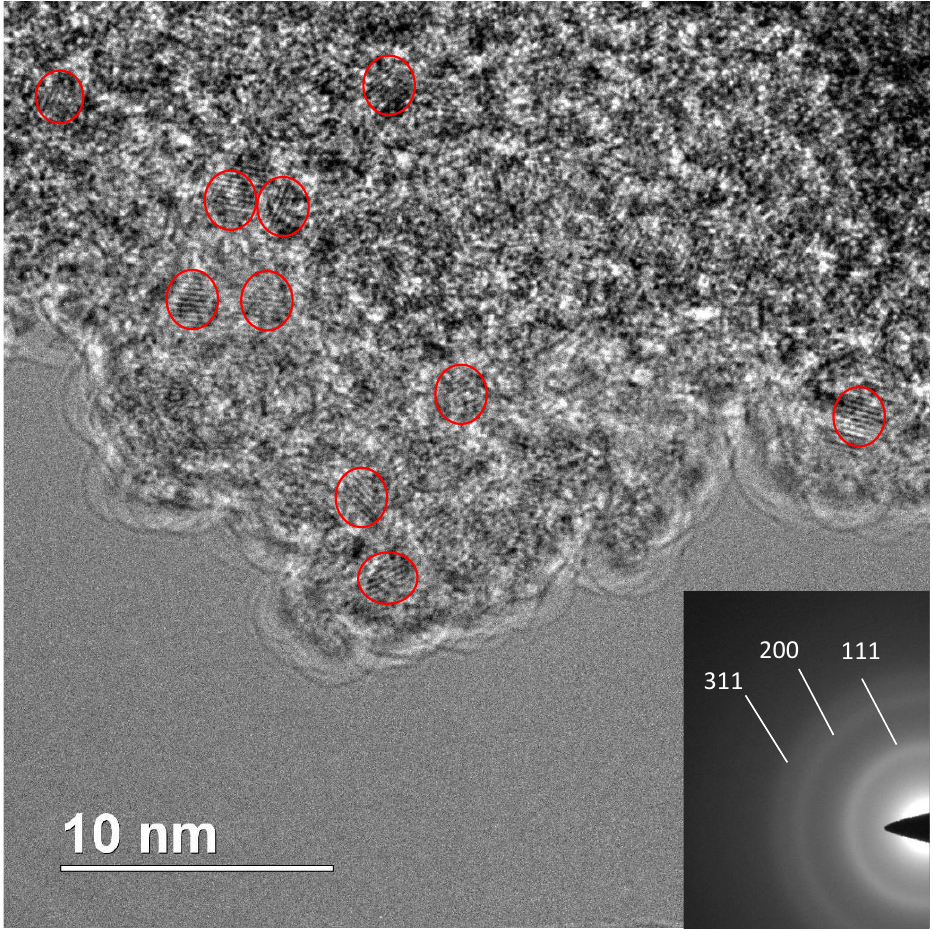}
\caption{TEM micrograph of the ``1.4 nm'' nanodiamonds synthesized at 800 $^\circ$C. Insert: electron diffraction pattern of nanodiamonds.}
\label{f:tem1}
\end{figure}

\begin{figure}
\includegraphics[width=\columnwidth]{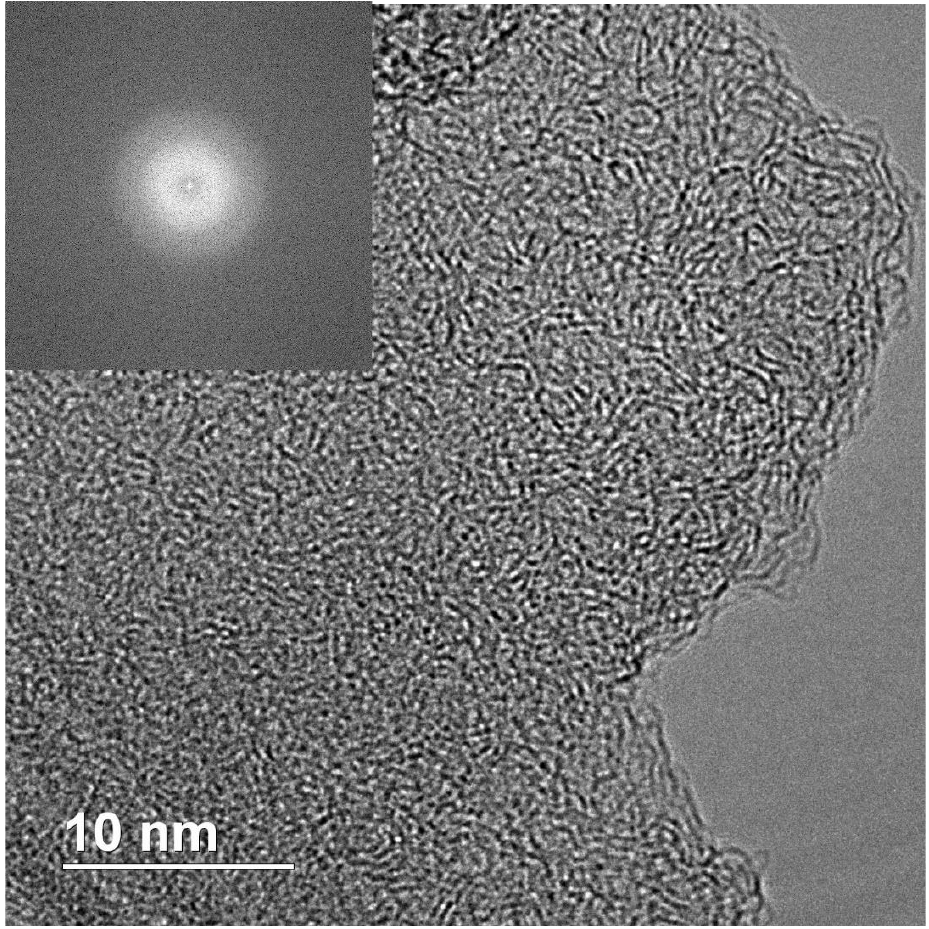}
\caption{TEM micrograph of the annealed ``1.4 nm'' nanodiamonds. Insert: fast Fourier transform calculated from the image of the amorphous carbon formed.}
\label{f:tem2}
\end{figure}

Figs.~\ref{f:7}, \ref{f:8}, \ref{f:tem1}, \ref{f:tem2} show TEM images of nanodiamonds with average size 3.4 nm and 1.4 nm before (Fig.~\ref{f:7},  \ref{f:tem1}) and after vacuum annealing (Fig.~\ref{f:8}, \ref{f:tem2}) at 1150 $^\circ$C.  Note that small as-synthesized nanocrystals (Fig.~\ref{f:tem1}) are covered with rather thick non-diamond shell, whereas products of their graphitization are  represented by poorly crystallized carbon. In contrast to annealed nanodiamonds of the 3.4 nm crystal size, no diamond crystals were detected in the graphitized material of the smallest nanodiamond sample by TEM (Fig.~\ref{f:tem2}). The crystal sizes of as-synthesized nanodiamonds measured directly on micrographs correspond to the average crystal size estimated by Scherrer's formula for peak (111). Obtaining good TEM images for statistical measurements of small nanodiamond is quite a difficult task, especially for nanodiamonds with a size smaller than 2 nm. For instance, in the paper \cite{fang:jn13}, authors stress that nanodiamonds were detected by X-Ray and electron diffraction methods but were not found by TEM even with at high magnification. In our case, in Fig.~\ref{f:tem1} with the 1.4 nm  nanodiamonds (TEM) we detect the smallest crystals of 0.8 nm and larger crystals of 1.8 nm. So, the average crystal size 1.4 nm, calculated by using Scherrer's formula on 111 peak falls in the 0.8-1.8 nm range quite well. In Fig.~\ref{f:7}, sizes of crystals vary from about 2 to 4 nm, and the average crystal size 3.4 nm falls inside the range. Statistics performed on TEM images can give somewhat different value of the average size purely due to difficulties in precise determination of a size of grains with nondiamond shell and because of their aggregation smearing boundaries of the grains.  Because of this, statistics on our TEM images will not give reliable values. 

We should note that our samples (despite the very broad diffraction peaks observed in X-ray diffraction on ultradisperse samples, see Fig.~\ref{f:2}) remains crystalline. It has been shown by scanning transmission electron microscopy that nanodiamonds stay crystalline down to 1 nm \cite{stehlik:jpcc15}. This is in perfect agreement with the XRD and Raman spectroscopy data provided here which both show clear signatures of diamond crystalline lattice down to 1.2 nm. Beside that the Raman spectra with diffuse lines are characteristic for nanodiamonds \cite{ferrari:ptrsa04,korepanov:c17}, it does not necessarily mean their amorphization.

%%%%%%%%%%%%%%%%%%%%%%%%%%%%%%%%%%%%%%%%%%
\section{Discussion}

\subsection{Synthesis mechanism}

Firstly, we would like to discuss possible mechanism leading to easy formation of nanodiamonds from halogenated adamantanes in comparison to pure adamantanes. As it was shown before for the case of PVC molecules (in fact short fragments of PVC molecules) \cite{kondrin:cnm21} the reaction involves several steps of pyrolysis accompanied by dehydrohalogenation of initial molecules, their polymerization and further halogenation of 2D/3D polymers. This promotes subsequent dehydrohalogenation and even higher degree of polymerization. The peculiar property of this reaction of nanodiamond formation is that in the ideal case it does not require breakage of C-C bonds (only C-H and C-Cl ones) and thus it can proceed at comparatively low temperature. However, it turns out that situation with adamantane is different. As it was mentioned before, the trial and error attempts lead us to conclusion that it is impossible to construct diamond crystal lattice with adamantane carbon backbones without breaking them. So, it is pertinent to pose a question which reaction involves the smallest possible number of breaking C-C bonds? We argue that the crucial step of this reaction is the formation of one diamantane molecule from two adamantanes:

\begin{multline}
2~C_{10}H_{15}Cl (chloroadmantane) \to C_{14}H_{20} (diamantane)\\
+ C_6H_{10}Cl_2 (cyclohexane~dichloride)
\label{ad_to_dia}
\end{multline} 
It requires breaking of only 3 C-C bonds per two adamantane molecules and reconstruction of C-H and C-Cl bonds. It seems that it is the rupture of a strong C-C bond that hinders the kinetics of this reaction and requires higher temperatures in comparison to the PVC case, which involves only dehydrohalogenation-mediated pyrolysis \cite{kondrin:cnm21}.

As discussed in Section 3.1, diamond lattice can be paved in 3D by backbones of diamantane molecules. It was previously shown that diamond lattice can be divided  into two interpenetrating networks \cite{kondrin:pccp15,kondrin:itmo16,kondrin:acb16} of carbon atoms. Each of these networks can be regarded as the carbon backbone of crystal polymeric hydrocarbon named diamond monohydride. On the other hand, diamond monohydride can be regarded as a result of polymerization of diamantane molecules with formation of 6 intermolecular C-C bonds between each diamantane molecule. These observations prove that the diamond lattice can be in principle built by pyrolysis of diamantane molecules and their subsequent polymerization. We should mention that mechanism of adamantane and diamantane molecule pyrolysis with formation of nanodiamond proposed in Ref.~\cite{park:sa20} is not viable because it involves creation of oligomers of adamantane and diamantane molecules including fragments of ordinary (cubic) diamond and its hexagonal counterpart (lonsdeleite). This would result in formation of heavily twinned nanodiamond crystals, which are not observed in experiments. Moreover, if we delve deeper into reaction Eq.~(\ref{ad_to_dia}), we can compare the predicted and experimentally observed (see Fig.~\ref{f:3})mass losses during diamond synthesis. We assume that cyclohexane chloride is volatile under the synthesis conditions and once created it leaves the reaction cell and does not participate in the diamond formation. Consequently, the ratio of carbon atoms remained in the carbonaceous material and removed from the system is 7/3=2.33. As follows from Fig.~\ref{f:3} the high-temperature asymptotic of this ratio in the experiment is 6.5/3.5=1.85. If we calculate the mass loss taking into account that the diamantane undergoes further pyrolysis and loses hydrogens the agreement with experiment will be even better and the mass loss will be equal to approximately  45\% (the high-temperature asymptote in Fig.~\ref{f:3} is about 54\%). So, we believe that the reaction Eq.~(\ref{ad_to_dia}) reproduces crucial features of the experiment. Although, we regard this mechanism as primary which accounts for a large part of mass loss observed in experiment, we can not exclude further cracking of adamantane molecules (especially at high temperatures) which would include formation of methane like this:

\begin{multline}
C_{10}H_{15}Cl (chloroadmantane) \to CH_4+ 3C \\
+C_6H_{11}Cl (cyclohexane~chloride)
\end{multline} 

This secondary mechanism produces 75 \% of mass loss (cyclohexane and methane are volatile products and are removed from the reaction cell), but it can account for the difference between mass loss predicted by the reaction described by Eq.~\ref{ad_to_dia} and observed in experiment. However, we believe that only small part of diamond is produced by this route, because this reaction requires breaking of larger number of C-C bonds so it proceeds at  higher temperatures in comparison to the reaction described by Eq.~\ref{ad_to_dia}. The similar consideration is applicable to the reaction which is shown in red in Fig.~\ref{f:3}. Although it does describe well the high-temperature asymptotic of mass loss observed in experiments it involves even larger number of C-C bond breaking. So, we can expect that it is possible at even higher temperatures than the reactions considered before.
\subsection{Fano effect in IR and Raman spectra}

As it seen from the IR spectra the ND samples with a size larger than  2.6 nm exhibit striking "transmission windows" in the region near 1332 cm$^{-1}$, {\em i.e.} in the region of Raman diamond mode.  The authors of Ref.~\cite{kudryavtsev21} propose that this effect can be due to the Fano resonance of diamond Raman mode and some continuum of modes connected to the conductive states on the nanodiamond surface. It seems that this behavior is not limited to diamond as similar Fano resonances at the phonon frequency were observed in doped single-walled carbon nanotubes \cite{lapointe:prl12}.

\begin{figure}
\includegraphics[width=\columnwidth]{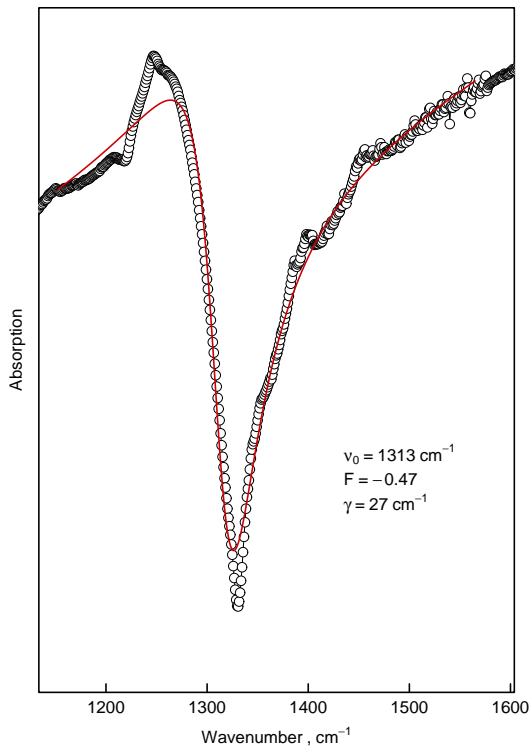}
\caption{The Fano lineshape fit (Eq.~\ref{eq:fano}, thick red curve) of the feature around 1300 cm$^{-1}$ ($\circ$) observed in the sample synthesized at 1300 $^\circ$C with background approximated by straight line. The nonlinear parameters of the fit are shown in the figure.}
\label{fano}
\end{figure}
The theory of the Fano resonances which is observed in many circumstances in the condensed matter physics is dated back to the work of Ugo Fano \cite{fano:pr61}. He proposed an asymmetric lineshape which, in contrast to symmetric Lorentz, Gauss and Voigt lineshapes, was suggested to describe frequency dependence of intensity $I(\nu)$ of experimentally observed resonances \cite{lukyanchuk:nm10,gallinet:prb11}:
\begin{equation}
I(\nu) \propto \frac{(F\gamma+(\nu-\nu_0))^2}{(\gamma^2+(\nu-\nu_0)^2)}
\label{eq:fano}
\end{equation}   
where $\nu_0$ and $\gamma$  are  parameters that describe the frequency and linewidth of the resonance, respectively; $F$ (called the Fano parameter) -- describes the asymmetry degree. The fit of the feature in the 8 nm sample synthesized at 1300 $^\circ$C by the Fano profile Eq.~(\ref{eq:fano}) is depicted in Fig.~\ref{fano}. This Figure unambiguously demonstrated the Fano like lineshape of this feature. However, the nature of the Fano resonance which give rise to this lineshape is discussable. The main prerequisite for observation of the Fano resonance is the strong coupling of narrow dark mode with the continuum (or at least with the band significantly broader than the dark mode) of optically active modes \cite{joe:ps06,gallinet:prb11,limonov:np17}. We should keep in mind that the diamond's Raman 1332 cm$^{-1}$ mode is dark in IR experiments, unless the inversion symmetry of diamond crystal is lost due to defects or some other reasons \cite{lax:pr55,birman74}. However, we consider this option as highly improbable for the pure nanodiamonds presented here.

It should be mentioned that the Fano resonances are observed in the total light scattering by nanoshells \cite{lukyanchuk:nm10,gu:n21}, {\em i.e.}  nanoparticles  consisting of a dielectric core surrounded by a metallic shell (usually made of silver or gold). Consequently, for nanodiamonds one could expect observation of Fano resonances in IR spectra provided that one can create conductive wrapping film around it. Kudryavtsev {\em et. al.} \cite{kudryavtsev21} suggested that formation of a surface conducting layer ( {\em i.e.} a film) is possible due to the well-known phenomenon of surface p-type conductivity observed in hydrogenated diamond in contact with water \cite{maier:prl00} and/or oxygen \cite{chakrapani:s07}. This phenomenon heavily relies on hydrogenation of diamond surfaces which leads to negative electron affinity (by about 1 eV below conduction band minimum) and on the presence of atmospheric surface adsorbates acting as acceptors for electrons from the diamond valence band. This results in band bending in diamond (by about 4 eV). Both factors lead to formation of subsurface conduction layer  caused by the rise of valence band maximum above the Fermi level in the diamond bulk. Still, we stress that this mechanism heavily depends on thermal diffusivity of electrons, so the conductive layer can be rather thick on the nanometer scale.

However, there are two objections in immediate applicability of this model for explanation of Fano resonances in FTIR spectra of nanodiamonds grown by HPHT techniques from chloroadamantane. First of all, it was reported that the subsurface conductive layer could be relatively thick (upto 30 nanometers in thickness \cite{maier:prl00}) so the total nanodiamond particle could be considered as conductive nanosphere. According to Ref.~\cite{lukyanchuk:nm10} such a nanosphere does not demonstrate Fano resonances in total light scattering (only if measured by differential technique, {\em e.g.} radar back scattering). However, the recent review seems to propose an existence of Fano resonances in conductive particles \cite{tribelsky:pu22}. So, for observation of Fano resonances in IR spectra in nanodiamonds one would  need a thinner conductive layer. The second objection, which is supported by our experimental findings, is no significant trace of water in our samples, as confirmed by IR spectroscopy. Therefore, another mechanism for appearance of a  conductive layer on the nanodiamond surface could be involved/envisioned.

\begin{figure}
\includegraphics[width=\columnwidth]{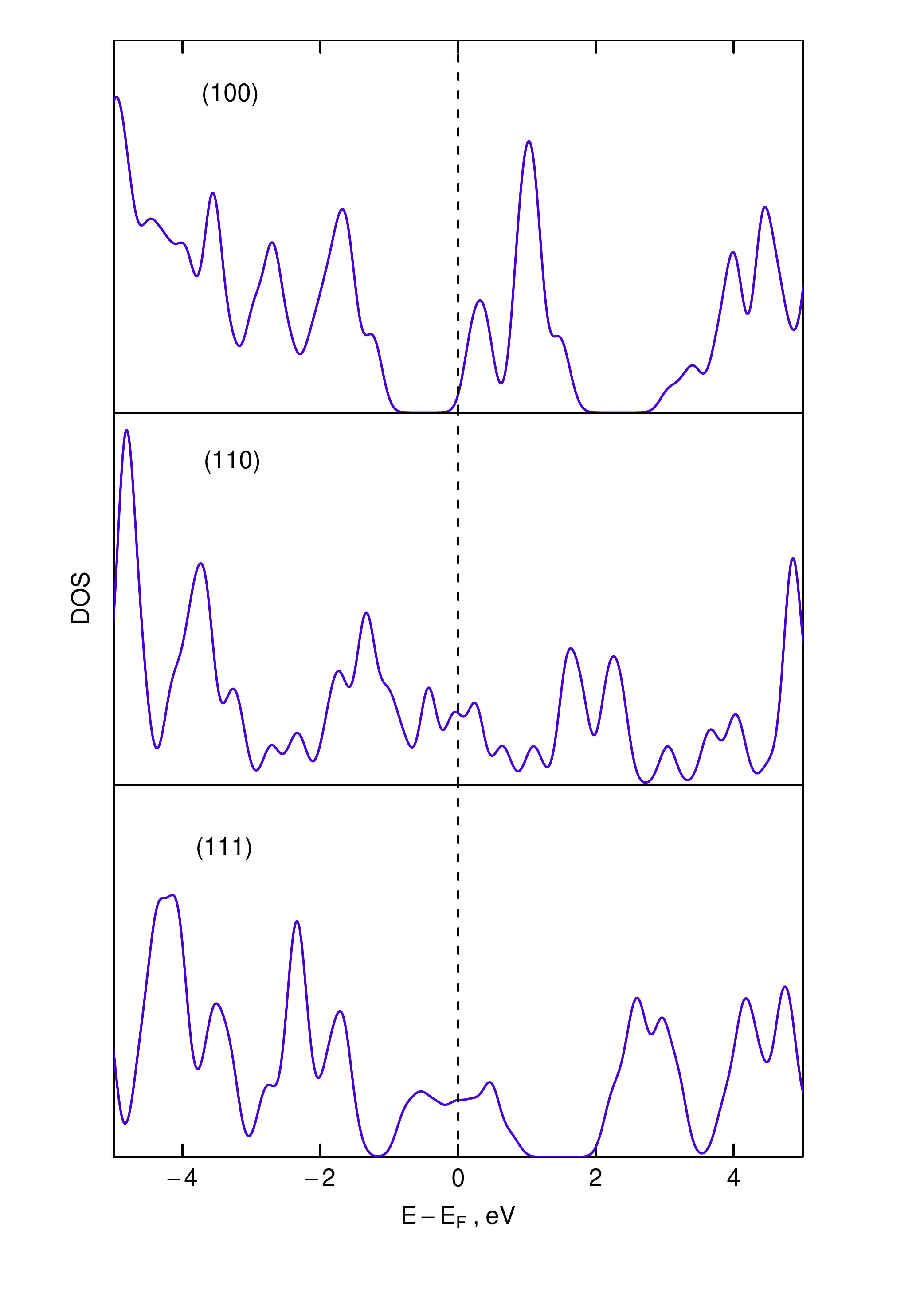}
\caption{The electron density of state calculated for three ultrathin diamond films in different orientation (the orientation of films is shown as labels in respective panels). The calculated Fermi energy is taken as zero.}
\label{dos}
\end{figure}

We propose that the surface conductive layer may arise due to surface reconstruction of dangling carbon bonds which could lead to formation of monoatomic thick conductive carbon layer. We hypothesize that for nanodiamonds with diameter below 5 nm the surface hydrogenation is not dominant  but the surface can be described as bare but reconstructed one. This assumption can explain the apparent dilatation of nanodiamonds observed in X-ray diffraction experiments \cite{ekimov:drm20}. This effect was explained by the strain exerted by the reconstructed surface on the ultra-small nanodiamonds \cite{ekimov:drm20,ekimov:c21}. In fact, for (110) surface these sp$^2$ hybridized carbon chains occur naturally, for (111) surfaces they appear due to 2$\times$1 reconstruction \cite{delapierre:mp14,ekimov:c21}, and only for (100) surface there is no conductive layer. In the latter case even 2$\times$1 reconstruction leads to formation of isolated sp$^2$-hybridized dimers. This observation can be proved by plotting the electron density of states for these three planes shown in the Fig.~\ref{dos}. For calculation of electron density of states we used ultrathin diamond films previously used for calculation of lateral expansion and elastic properties \cite{ekimov:drm20,ekimov:c21}. All calculations were carried out in GGA approximation which leads to the decreased value of bandgap ($\approx$ 4.2 eV). As it can be seen from Fig.~\ref{dos} only for (111)-2$\times$1 and (110) surfaces the density of state has appreciable value at the Fermi level. For reconstructed (100) surface the Fermi level is located just below the energy range where electron density of states reaches appreciable value. For macroscopic diamond particles this monoatomic conductive layer is difficult to identify but it can produce a significant impact on conductivity of nanoscale samples where the ratio of surface to the bulk atoms is extremely high.

This bears important implications for explanation of evolution of IR spectra in vicinity of 1332 cm$^{-1}$ mode {\em vs.}  the size of nanodiamond particles. According to theoretical investigations \cite{barnard:drm03} the existence of nanodiamond with (111) faces with diameter below 2 nm is impossible due to graphitization and exfoliation of the (111) surfaces. Thus, in the size range below 2 nm a drastic change in ratio of nanodiamond particles with conductive ((111) and (110)) and nonconductive (100) faces should occur. Consequently, this change should result in significant attenuation and eventually disappearance of Fano resonance observed in IR spectra in vicinity of diamond Raman mode which is indeed observed in experiment for samples with diameter between 1.4 and 2.6 nm (see Fig.~\ref{f:4}). We can conjecture that the remnants of continuum of IR active modes visible in samples with size below 2.6 nm as a ``forest'' of absorption peaks around 1300 cm$^-1$ are more or less the same modes that lead (after appropriate coupling with dark Raman mode) to the appearance of the Fano resonance observed in the samples with larger crystallites size.

This hypothesis, however, appears to contradict experimental observation of C-H vibrations abundance. We argue that the signals of the C-H functional groups originate from the isolated patches of hydrogenated diamond surface. They contribute to IR and Raman spectra but do not break the contiguity of the conductive sp$^2$-hybridized carbon layer. We conclude here that  both effects (transfer doping mechanism as well as the monoatomic conductive reconstruction) could contribute to the observed Fano resonance effect in IR absorption.

Another possible mechanism of conductive surface states could be associated with trans-polyacetylene fragments adsorbed on the surface of nanodiamonds. This possibility is suggested by the presence of peaks at 1150, 1250 and 1450 cm$^{-1}$ visible in the Raman spectra (Fig.~\ref{f:5}) which are usually ascribed to trans-polyacetylene vibrations. Still as we explain below there is some difficulties with this ascription in our case.

The well-established fact is the appearance of lines at 1150, 1250 and 1450 cm$^{-1}$ in the Raman spectrum of CVD nanodiamonds and they are absent in detonation nanodiamonds.  The nature of these lines was explained by the appearance of transpolyacetylene segments at grain/film surface during deposition \cite{lopez:prl96,ferrari:prb01}. At the same time, Lopez-Rios et al. \cite{lopez:prl96} notes the unnatural strengthening of the lines for trans polyacetylene after additional annealing of diamond films at 850 $^\circ$C in vacuum, which cannot be rationally explained, because it is known that transpolyacetylene decomposes at normal pressure at a temperature of 420 $^\circ$C \cite{ito:jps75}. In CVD process, diamond nanofilms were obtained at temperatures of about 700-800 $^\circ$C, where trans polyacetylene should be completely decomposed. The observation of lines 1150, 1250 and 1450 cm$^-1$ lines in the Raman spectra of our nanodiamonds synthesized at 1300 $^\circ$C raises even more doubts that the lines belong to ordinary trans-polyacetylene. It is known that except methane, characteristic temperature interval of hydrocarbon decomposition at 8 GPa is 600-800 $^\circ$C \cite{davydov:c04,kondrin:cec17}.  In this context, an abnormal increase in the thermal stability of transpolyacetylene from 420 $^\circ$C to 1300 $^\circ$C looks unreal.

In fact, the modes at 1150 and 1450 cm$^{-1}$ in the Raman spectrum of CVD diamonds are confidently associated with the C=C  stretching \cite{lopez:prl96,ferrari:prb01}, while their attribution to transpolyacetylene was rather tentative and has no direct proof. Taking into account evolution of spectra with temperature synthesis, we can rather agree with Ref.~\cite{lopez:prl96}  that the line at 1250 cm$^{-1}$ corresponds to the maximum in the phonon DOS in diamond rather than to the assignment of the line to transpolyacetylene \cite{ferrari:prb01}. Thus, we can say with certain confidence that the appearance of the lines at 1150 and 1450 cm$^{-1}$ in spectra of our nanodiamonds is associated with the presence of a double C=C bonds.

For the Raman mode the situation with the Fano effect is different. In the Raman measurements the narrow mode is bright (observable) as is. So, the Fano effect (interference of narrow mode with the continuum) leads only to small asymmetric contribution to the intense Raman line. This attenuation is observed in Ref.~\cite{kudryavtsev21} and in our measurements (see Fig.~\ref{f:5}). We can conclude that although the Fano effect is also observable in the Raman measurements, it is not as spectacular as in IR experiments. 

\subsection{Electrical conductivity measurements}

 To further elucidate the origin of the Fano resonance we perfomed electrical conductivity and FTIR measurements. Electrical conductivity of two case samples with (8 nm) and without (1.2 nm) IR transmission window at 1330 cm$^{-1}$ was measured right after deposition on IDE sensor (dropcasting and short drying at 70 $^\circ$C until complete evaporation of toluene). Then the sensor with the deposited sample was annealed at 250 $^\circ$C for 1 h and conductivity was recorded as a function of time for next 24 h. The annealing step was employed in order to possibly distinguish between two assumed conductivity mechanism, i.e. transfer doping and a conductive surface structure. 

The 1.2 nm sample was non-conductive (conductivity was lower than 10$^{-12}$ S/cm, i.e. below the measurement setup limit) after deposition and toluene evaporation at 70 $^\circ$C. As soon as the temperature increased to 250 $^\circ$C the conductivity increased to 10$^{-7}$ S/cm but as soon as the sample cooled down the conductivity again dropped below 10$^{-12}$ S/cm. 

In contrast, the 8 nm sample was conductive immediately after the deposition, the conductivity was 2.10$^{-5}$ S/cm. 

\begin{figure*}
\includegraphics[width=\textwidth]{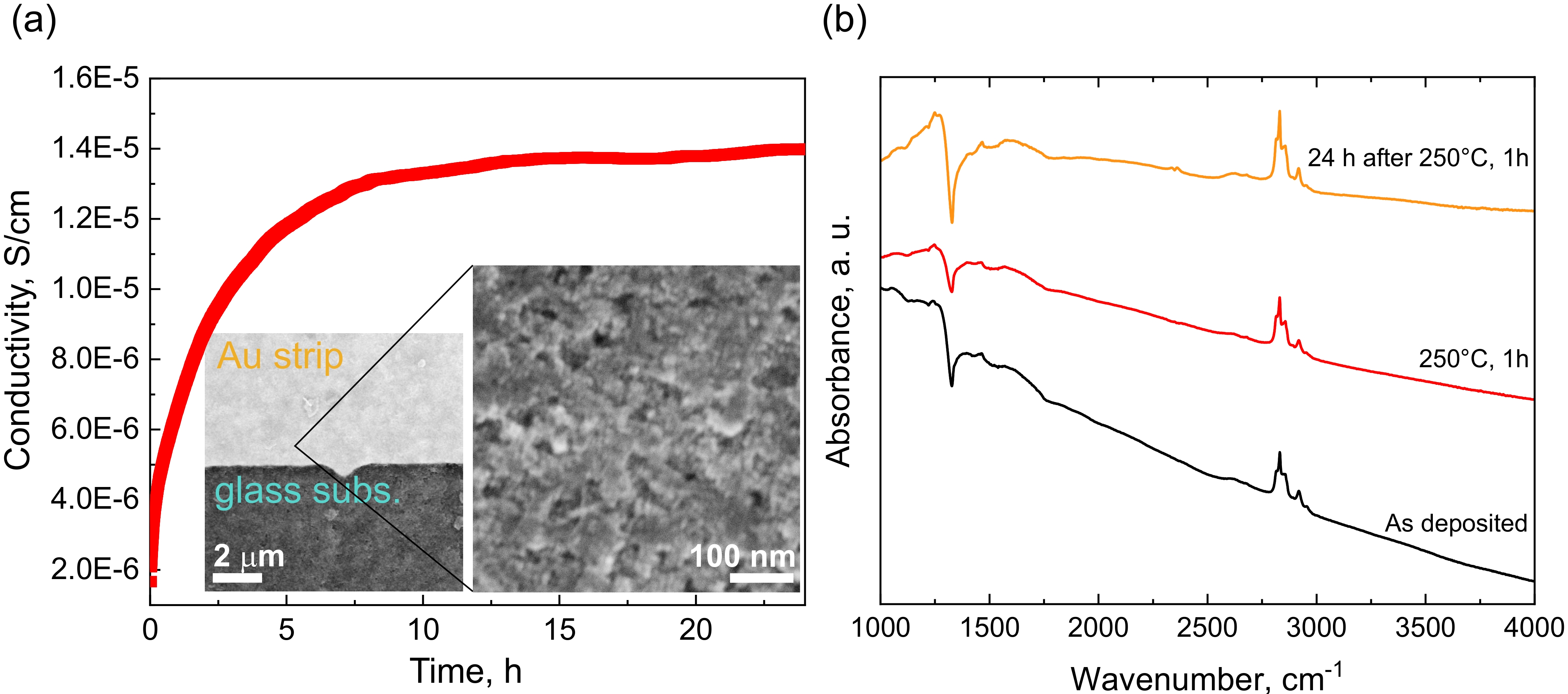}
\caption{Conductivity as function of time of the 8 nm sample after 1 h annealing at 250 $^\circ$C. The inset shows continuous nanodiamond layer deposited on the IDE sensor (a). FTIR spectra of the 8 nm sample recorded after deposition on Au mirror (black), after 1 h annealing at 250 $^\circ$C (red), and after 24 h time period after the 1 h annealing at 250 $^\circ$C (yellow). }
\label{f:cond}
\end{figure*}

After 1h annealing at 250 $^\circ$C the conductivity was approx. one order of magnitude lower (1.7 10$^{-6}$ S/cm) and increased again to 1.4 10$^{-5}$ S/cm after 24 h, i.e. close to the initial value as shown in Fig.~\ref{f:cond}~a). SEM image of the continuous nanodiamond layer as deposited on the Au-glass interface of the sensor is also shown in the Fig.~\ref{f:cond}~a).

The observed behavior somewhat resembles behavior reported for hydrogenated bulk diamond surface where transfer doping takes place. However, there are two striking differences: i) the 8 nm sample remain well-conductive after annealing at 250 $^\circ$C for 1h, and ii) only one order of magnitude drop of the conductivity occurs after the annealing. Hydrogenated diamond surface exhibiting the transfer doping effect loses its conductivity above 200$^\circ$C and the conductivity change is as high as 5 orders of magnitude \cite{maier:prl00}.

 Thus, the electrical conductivity data seem to establish the link between the IR transmission window and the presence of free carriers, i.e. conductivity as well as suggest that both mechanisms are, to an unknown extent, involved. To further confirm the link between the IR transmission window and the conductivity, we recorded also FTIR spectra of the as deposited (black), annealed at 250 $^\circ$C for 1h (red) and after 24 h at ambient (yellow) shown in Fig.~\ref{f:cond}~b). Overall, the spectra show negligible effect of the annealing (250 $^\circ$C, 1h) on the 8 nm sample surface chemistry, {\em i.e.} no oxidation occurs. The only noticeable change is difference in the transmission window intensity between the annealed state and the two ambient states. The lower intensity of the Fano resonance at 1330 cm$^{-1}$ of the annealed sample clearly correlates with the decrease of conductivity and thus confirm the link between the IR transmission window and the conductivity in the studied nanodiamond samples.

Although we provide solid arguments, the results should be still considered as preliminary and further work is needed to elucidate contribution and details of both conductivity mechanisms.   

\subsection{Thermal annealing}

In electron diffraction pattern of the annealed sample (Fig.~\ref{f:8}), apart from reflections of diamond, a wide halo corresponding to the interplanar distance of 0.35 nm of graphite-like carbon is detected.  Similar to graphitized DND \cite{butenko:jap00}, the graphitized products of our sample are represented by onion-like forms of carbon with sizes comparable to the sizes of the initial nanodiamonds.   At the same time, nanocrystals of the original size are present in the sample, which is an unexpected result from the point of view of idealized mechanism of the surface graphitization of DND \cite{qiao:sm06,kuznetsov:jap99,tomita:jcp01,zhao:drm02}. In spite of nanodiamond aggregation, the mechanism suggests that the  crystals graphitize uniformly as a result of surface graphitization activated by residual oxygen (and other reactive volatiles) in vacuum or inert atmosphere \cite{kuznetsov05}. This contradiction is lifted if we assume that the penetration of residual impurities along the intergrain boundaries into the aggregates is impeded and/or these reactants are consumed during graphitization of surficial grains of the aggregate. In this case, the process of graphitization of nanodiamonds should be considered as essentially heterogeneous with preservation of the original structure of nanocrystals under inside dense aggregates. Note also that our samples are composed of nanocrystals densely packed into rather large transparent pieces (Fig.~\ref{f:1}).

Diamond crystallites were not observed by TEM or electron diffraction in annealed sample with initial grain size of 1.4 nm (Fig.~\ref{f:tem2}). Considering corresponding data of X-ray phase analysis, we conclude that nanodiamonds were completely transformed to highly disordered sp$^2$-carbon. In contrast to the heterogeneous graphitization of larger nanodiamonds, the complete destruction of the 1.4 nm nanodiamonds indicates the bulk character of the transformation. Peculiar morphology of the annealed product, in which the presence of onion-like carbon is not typical, also supports the bulk graphitization mechanism. Thus, the study of graphitization of nanodiamonds with different sizes does not confirm increase in thermal stability of nanodiamonds less than 2 nm, which could have been expected on basis of theoretical predictions and experimental studies \cite{zhao:drm02, sun:jpcl14, barnard:jcp03, wang:acie05}. Our investigation suggests also that in contrast to the surface graphitization of the 3.4 nm nanodiamonds, bulk graphitization take place for smaller nanodiamonds with a size of 1.4 nm.

%%%%%%%%%%%%%%%%%%%%%%%%%%%%%%%%%%%%%%%%%%
\section{Conclusions}

We performed investigation of optical properties and structural stability of ultra-small nanodiamonds synthesized from chloroadamantane under constant pressure of 8-9 GPa and varying temperature. Based on the temperature dependent mass loss and XRD data we suggested possible mechanism of the nanodiamond formation above 500 $^\circ$C where the crucial steps are formation of diamantane molecules as intermediate step and their subsequent polymerization triggered by dehydrohalogenation of diamantane molecules. XRD and TEM study of graphitization of nanodiamonds with different crystal sizes does not confirm theoretically predicted increase in thermal stability of nanodiamonds smaller than 2 nm. IR spectra of nanodiamonds revealed a "transmission window" at $\approx$ 1330 cm$^{-1}$ {\em i.e.} at the phonon frequency of diamond. We attributed this feature to the Fano resonance of diamond Raman mode with continuum of free electronic states originated from  conductive surface states. Due to several objections to transfer doping mechanism we assume that  these conductive states may be linked to the reconstruction of carbon atoms on the surface of nanodiamonds. 

The disappearance of the 1330 cm$^{-1}$ transmission window is observed for nanodiamonds with size smaller than 2 nm. We tentatively propose that this may be due to the changes in morphology of nanodiamonds and  corresponding changes in ratio of conductive (111) and (110) faces to non-conductive (100). 

\section*{Acknowledgements}
This work is supported by Russian Foundation for Basic Research (Grant No. 20-52-26017), by Czech Science Foundation (GACR) (Grant  No. 21-12567J) and by CzechNanoLab research infrastructure supported by the Ministry of Education, Youth and Sports of the Czech Republic (LM2018110)

XRD, Raman and IR spectra were obtained using equipment of Center of shared use of IPCE RAS. The electron microscopy was performed using the equipment of the Shared Research Center "Structural diagnostics of materials" and partially supported by the Ministry of Science and Higher Education within the state assignment FSRC ``Crystallography and Photonics'' RAS. We acknowledge Ondrej Szabo for the conductivity experiments and Jitka Libertinova for SEM work.

%\bibliography{fano}

\end{document}